\documentclass[twocolumn,twoside,slac_two]{revtex4}
\usepackage{graphicx}
\usepackage{fancyhdr}
\begin{document}

\lefthyphenmin=5
\righthyphenmin=5

\def\pbar{$\overline{p}$}               %pbar
\def\tbar{$\overline{t}$}               %tbar
\def\bbar{$\overline{b}$}               %bbar
\def\pizero{$\pi^0$}                    %pizero
\def\jpsi{$J/\psi$}                     %J/psi
\def\ppbar{$p\overline{p} $}            %ppbar
\def\pbarp{$\overline{p}p $}            %pbarp
\def\qqbar{$q\overline{q}$}             %qqbar
\def\ttbar{$t\overline{t}$}             %ttbar
\def\bbbar{$b\overline{b}$}             %bbbar
\def\epm{$e^+e^-$}                      %e+e-
\def\wino{$\widetilde W$}               %Wino
\def\zino{$\widetilde Z$}               %Zino
\def\squark{$\widetilde q$}             %squark
\def\gluino{$\widetilde g$}             %gluino
\def\ttbarH{$t\overline{t}H$}           %ttbarH
\def\ttbarHbbbar{$t\overline{t}H(\to b\overline{b})$}
\def\Hbbbar{$H(\to b\overline{b})$}
%
% kinematics, physics quantities, etc.
\def\pt{$p_T$}                          %pT
\def\et{$E_T$}                          %ET
\def\met{\mbox{${\hbox{$E$\kern-0.6em\lower-.1ex\hbox{/}}}_T$}} %missing ET
\def\htran{$H_T$}                       %HT
\def\aplan{$\cal{A}$}                   %aplanarity
\def\iso{$\cal{I}$}                     %isolation variable
\def\remu{${\cal{R}}_{e\mu}$}           %distance between e and mu in eta-phi
\def\rmu{$\Delta\cal{R}_{\mu}$}         %distance between mu and jet cone axis
\def\sigtot{$\sigma_{\rm tot}$}         %sigma total
\def\sigtop{$\sigma_{t \overline{t}}$}  %sigma_ttbar
\def\njet{$N_{\rm jet}$}                %N_jet
\def\sinthw{sin$^2 \theta_W$}           %sin^2 th_W
\def\alphas{$\alpha_{\scriptscriptstyle S}$}                %alpha_s
\def\alphaem{$\alpha_{\scriptscriptstyle{\rm EM}}$}         %alpha_EM
%
% units
\def\ipb{pb$^{-1}$}                     %inverse picobarns
\def\gevcc{GeV/c$^2$}                   %GeV/c^2
\def\gevc{GeV/c}                        %GeV/c
\def\gev{GeV}                           %GeV
\def\deg{$^\circ$}                      %degree sign
%
% luminosity stuff
\def\lum{$\cal{L}$}                     %luminosity
\def\lumint{$\int\! {\cal{L}} dt$}      %integrated luminosity
\def\lumunits{cm$^{-2}$s$^{-1}$}        %luminosity units
%
% masses
\def\mt{$m_t$}                          %m_top
\def\mb{$m_b$}                          %m_bottom
\def\mw{$M_W$}                          %M_W
\def\mz{$M_Z$}                          %M_Z
%
% misc.
\def\D0{D\O}                            %D0
\def\etal{{\sl et al.}}                 %et al. - no preceeding comma
\def\vs{{\sl vs.}}                      %vs.
\def\d0draft{}
\def\d0{D\O}
\def\wjets{\mbox{$W +$ Jets}}
\def\pbarp{\mbox{$\overline{p}p$}}
\newcommand{\comphep}   {\sc comphep}
\newcommand{\singletop} {\sc singletop}
\newcommand{\pythia}    {\sc pythia}
\newcommand{\alpgen}    {\sc alpgen}
\newcommand{\mcfm}      {\sc mcfm}
\newcommand{\geant}     {\sc geant}
\newcommand{\madgraph}    {\sc{madgraph}}

\title{Observation of Single Top Quark Production}


\author{Cecilia E. Gerber \footnote{\it on behalf of the D0 Collaboration}}
\affiliation{Department of Physics, University of Illinois at Chicago, Chicago, IL 60607, USA}

\begin{abstract}
I report on the observation of electroweak production of single top quarks 
in \ppbar\ collisions at $\sqrt{s}=1.96\;\rm TeV$ using $2.3\;\rm fb^{-1}$ 
of data collected 
with the D0 detector at the
Fermilab Tevatron Collider. Using events containing an isolated electron or muon, 
missing transverse energy, two, three or four jets, with one or two 
of them identified as originating
from the fragmentation of a $b$ quark, the measured cross section for the 
process
$p\overline{p} \to tb+X,~tqb+X$ 
is $3.94 \pm 0.88$~pb (for 
a top quark mass of $170\;\rm GeV$). The
probability to measure a cross section at this value or higher in the
absence of signal is $2.5\times10^{-7}$, corresponding to a
5.0~standard deviation significance. Using the same dataset, the measured cross 
sections
for the $t$- and the $s$-channel processes when 
determined simultaneously with no assumption on their relative production 
rate are
$3.14^{+0.94}_{-0.80}$~pb and $1.05 \pm 0.81$~pb respectively,
consistent with standard model expectations. The measured $t$-channel 
cross section has a significance of 4.8 standard deviations, representing the 
first evidence for the production of an individual single top process to be 
detected. 

\end{abstract}

\maketitle

\thispagestyle{fancy}


%%%%%%%%%%%%%%%%%%%%%%%%%%%%%%%%%%
\section{Introduction}

The top quark's large mass, by far the heaviest fundamental particle known, 
makes it a unique probe of physics at the natural electroweak scale. 
Although the top quark mass is consistent with other precision electroweak 
measurements within the framework of the standard model (SM), 
no fundamental explanation 
exists for the reasons why top quarks should be so massive. The large mass of the 
top quark arises from its large couplings to the symmetry breaking sector of the 
SM. Precision measurements of the top mass, width and couplings may 
therefore lead to a deeper understanding of electroweak symmetry breaking and 
the origin of mass. Such measurements are possible in part because the top quark's 
natural width of $1.4\;\rm GeV$ is much greater than the hadronization timescale set 
by $\Lambda_{\rm QCD}$, causing the top quark to decay to a real $W$ boson and a 
bottom quark before hadronization. The top quark can therefore be completely described 
by perturbative QCD, and studied as a bare quark. 

The SM predicts that top quarks are created via two 
independent production mechanisms in hadron colliders. 
The primary mode, in which a \ttbar\ pair is produced from a $gtt$ 
vertex via the strong interaction, was used by the D0 and CDF 
collaborations to establish the existence of the top quark in 
1995~\cite{top-obs-1995-cdf,top-obs-1995-d0}, 
and measure its mass. 
The second production mode of top quarks at hadron colliders is the 
electroweak production of a single top quark from a $Wtb$ vertex. 
The predicted cross section for single top production is about half 
that of \ttbar\ pairs, but the signal-to-background ratio is much worse; 
measurement of the single top quark production cross
section has therefore until recently been impeded by its low rate and difficult 
background environment compared to the top pair production.

Extracting a single top signature would be an independent confirmation of the top 
quark existence and a means to measure the Cabibbo-Kobayashi-Maskawa (CKM) matrix 
element $V_{tb}$, and study the $Wtb$ coupling. Measuring the single top quark production 
cross section is also crucial in understanding the backgrounds to searches for 
the Higgs Boson and new physics beyond-the-SM. In this paper I describe 
the analysis that led to the observation of single top production~\cite{d0_obs} by the D0
collaboration, and the measurement of the corresponding production cross sections. 

%%%%%%%%%%%%%%%%%%%%%%%%%%%%%%%%%%
\section{Single Top Quark Production}
\begin{figure}[!h!btp]
\vspace{0.1in}
\includegraphics[width=0.45\textwidth]{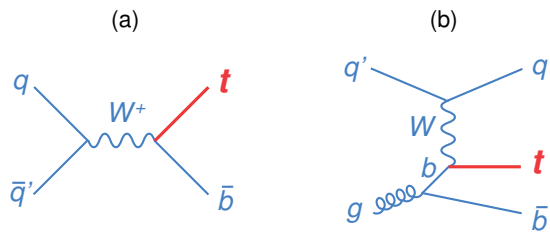}
\vspace{-0.1in}
\caption{Main tree-level Feynman diagrams for (a) $s$-channel and (b) $t$-channel 
single top quark production.}
\label{fig:feynman}
\end{figure}

In the SM, single top production at hadron colliders provides an 
opportunity to study the charged-current weak-interaction of the top quark. 
Figure~\ref{fig:feynman} shows representative Feynman diagrams for single top quark 
production at hadron colliders for $s$-channel (Fig. 1a), and $t$-channel production 
(Fig. 1b). A third process, usually called ``associate production'', 
in which the top 
quark is produced together with a $W$ boson, has negligible cross section at the 
Tevatron.

The SM predicts that the top quark decays almost exclusively to a 
$W$ boson and a bottom quark with $B(t \to Wb) \approx 1$. The rate for the process 
leads to a firm prediction for the top quark decay width $\Gamma_{t}$. 
A direct measurement of $\Gamma_{t}$ is of great importance, because the width would 
be affected by any non-expected decay modes of the top quark, whether they are 
observed or not. Unfortunately, $\Gamma_{t}$ cannot be directly measured in the 
\ttbar\ sample at hadron colliders, but its main component can be accessed 
through single top processes. If there are only three generations, the unitarity 
constrain of the CKM matrix implies that $|V_{tb}|$ is very close to unity. But, 
the presence of a heavy fourth generation quark with a large CKM coupling to the 
top quark could be consistent with large values of $B(t \to Wb)$, while resulting in 
an almost entirely unconstrained value for $|V_{tb}|$. A direct measurement of 
$|V_{tb}|$ can therefore explore the possibility of a fourth generation, and confirm 
that the top quark discovered at the Tevatron is indeed the SU(2) partner of the bottom 
quark.

As can be seen from the Feynman diagrams, 
the single top production cross section 
is proportional to $|V_{tb}|^2$. A measurement of the single top quark production 
cross section therefore provides the only known way to directly measure $|V_{tb}|$ at a 
hadron collider. 

%%%%%%%%%%%%%%%%%%%%%%%%%%%%%%%%%%
\section{Event Selection and Modeling}
The D0 detector~\cite{d0_nim}  is a
multi-purpose apparatus designed to study \ppbar\ collisions at high
energies. It consists of three major subsystems.
At the core of the detector, a
magnetized tracking system precisely records the trajectories of charged
particles and measures their transverse momenta. A hermetic,
finely-grained uranium and liquid argon calorimeter measures the energies of
electromagnetic and hadronic showers.
A muon spectrometer measures the momenta of muons.

The result presented in this document is based on $2.3\;\rm fb^{-1}$ of data recorded
using the D0 detector between 2002 and 2007. The data were collected with a logical OR of many trigger 
conditions that results in a fully efficient trigger selection for the single top signal.  
Events are selected 
containing exactly one isolated high $p_T$ electron or muon, missing 
transverse energy, and at least two jets, with at least one jet being identified as
originating from the fragmentation of a $b$ quark. The $\met$ is required not to be 
aligned with the direction of the lepton or the leading jet to limit the number of 
events originating from QCD multijet production entering our candidate samples. 
The data are divided into 24 mutually exclusive subsamples
to take advantage of the different signal:background ratios and dominant sources 
of background. The sample is divided based on the Tevatron running period 
(Run IIa or Run IIb), the lepton flavor 
($e$ or $\mu$), the jet multiplicity (2, 3 jets or 4 jets), and the number
of jets identified as originating from $b$ quarks (1 or 2 $b$-tags). In each
case, the leading $b$-tagged jet is required to have $p_T>20\;\rm GeV$. 
The efficiency of this selection is 
$(3.7 \pm 0.5)\%$ for $s$-channel and $(2.5 \pm 0.3)\%$ for $t$-channel single top production.

Single top signal events are modeled 
using the {\comphep}-based next-to-leading order
(NLO) Monte Carlo (MC) event generator
{\singletop}~\cite{singletop-mcgen}. The
{\singletop} generator is chosen as it preserves the spin information for the decay products of the top quark 
and resulting $W$~boson. {\pythia}~\cite{pythia} is used to model the
hadronization of any generated partons. We assume SM
production for the ratio of the $tb$ and $tqb$ cross sections.
The {\ttbar}, $W$+jets, and $Z$+jets backgrounds are simulated using
the {\alpgen} leading-log MC event generator~\cite{alpgen}, with 
{\pythia} used to model hadronization. The {\ttbar} background is
normalized to the predicted cross section for a top quark mass of $170\;\rm GeV$~\cite{ttbar-xsec}. 
The normalization of the $W/Z$+jets background is obtained by scaling the {\alpgen} cross sections 
by factors derived from calculations of
NLO effects~\cite{mcfm}. $Wb\bar{b}$ and $Wc\bar{c}$ are scaled by
1.47, and $Wcj$ by 1.38. $Zb\bar{b}$ and $Zc\bar{c}$ are scaled by
1.52 and 1.67. Diboson backgrounds are modeled using
{\pythia}. 

All MC events are passed through a {\geant}-based
simulation~\cite{geant} of the D0 detector. 
Small additional corrections are applied to all
reconstructed objects to improve the
agreement between collider data and simulation.
In particular, we correct mismodeling of the pseudorapidity of the jets, 
and the distance between the two leading jets in the $W$+jets
sample. 

The multijet background is modeled using collider data 
containing leptons that are not isolated. 
In the electron channel, the transverse momentum of the lepton is reweighted to properly match 
the shape of the background events passing the candidate selection. 
To increase the statistics in the muon channel, the jet
closest to the muon is removed and {\met} recalculated. 
Cuts on the total transverse energy of the event ($H_T$)
ensure that the multijet
background is a small (less than 5\%) contribution to the candidate sample. 
The overall normalization of the multijet and the total $W$+jets background is obtained 
by comparing the background expectation to collider data in three sensitive variables:
$p_T(\ell)$, {\met}, and the $W$~boson transverse mass. This normalization is done 
after subtracting from the data sample the contributions from the small backgrounds 
(\ttbar\ , $Z$+jets, and dibosons) separately for each
running period (Run IIa or Run IIb), leptonic channel ($e$ or $\mu$), and jet multiplicity bin 
(2, 3, or 4 jets). The normalization is performed before $b$-tagging, when the expected signal to background
ratio is on average S:B=1:260. 

The probability of the $b$-tagging algorithm to identify a jet as originating from a $b$-quark is 
measured in data containing jets and muons, and is parametrized as a function of jet flavor, $p_T$, and
$\eta$ with so-called tag-rate-functions. 
We find that after $b$-tagging, an additional empirical correction needs to be applied to 
the normalization of the $Wb\bar{b}$ and $Wc\bar{c}$ samples. 
The correction factor of $0.95 \pm 0.13$ is 
derived from the two-jet data sample and applied to all jet multiplicity bins; 
the overall $W$+jets normalization remains unchanged. After $b$-tagging the expected
S:B=1:21 for the sample with 1 $b$-tag and 1:15 for the sample with 2 $b$-tags.    

Using $2.3\;\rm fb^{-1}$ of data we selected 4,519 events, 
and expect $223\pm 30$ single top quark events. The summary of 
event yields as a function of jet multiplicity can be seen in Table~\ref{tab:event-yields}. 

\begin{table}[!h!btp]
\caption{Number of expected and observed events in
2.3~fb$^{-1}$ of D0 data. In the table, the event yields for $e$ and $\mu$, and 1 and 2 $b$-tagged 
jets have been combined.}
\label{tab:event-yields}
\begin{tabular}{|l|c|c|c|}\hline
Source & 2 jets & 3 jets & 4 jets \\ 
\hline
$tb$+$tqb$ signal         &   $139\pm 18$ &    $63 \pm  10$ &  $21 \pm  5$ \\
$W$+jets                  & $1,829 \pm 161$ &   $637 \pm  61$ & $180 \pm 18$ \\
$Z$+jets and dibosons     &   $229 \pm  38$ &    $85 \pm  17$ &  $26 \pm  7$ \\
\ttbar\ $\to \ell\ell$, $\ell$+jets
                          &   $222 \pm  35$ &   $436 \pm  66$ & $484 \pm 71$ \\
Multijets                 &   $196 \pm  50$ &    $73 \pm  17$ &  $30 \pm  6$ \\ 
Total prediction~~        & $2,615 \pm 192$ & $1,294 \pm 107$ & $742 \pm 80$ \\ \hline
Data                      & 2,579
                          & 1,216
			  & 724  \\    \hline
\end{tabular}
\end{table}

Systematic uncertainties are considered for all corrections 
applied to the background model. Most affect only the normalization,
but the corrections for the jet energy scale (JES), the
tag-rate functions (TRF), and the reweighting of the kinematic distributions 
in $W$+jets events modify in addition the shapes of the background
distributions. The
largest uncertainties come from JES and TRF, with smaller
contributions from MC statistics, the correction for jet-flavor
composition in $W$+jets events, and from the $W$+jets, multijets, and
{\ttbar} normalizations. The total uncertainty on the background is
(8--16)\% depending on the analysis channel. 

\begin{figure*}[!h!tbp]
\centering
\includegraphics[width=2.5in]{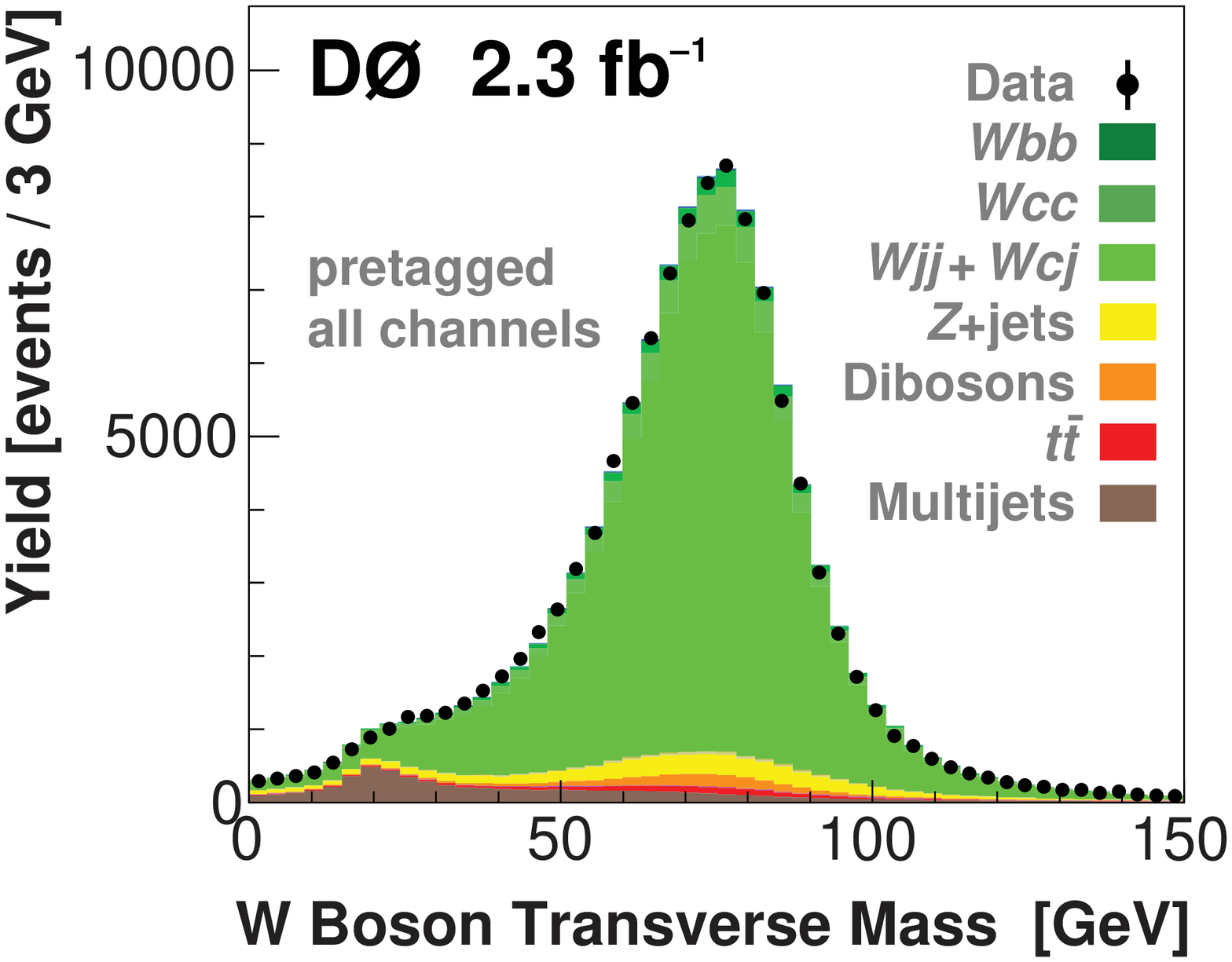}
\includegraphics[width=2.5in]{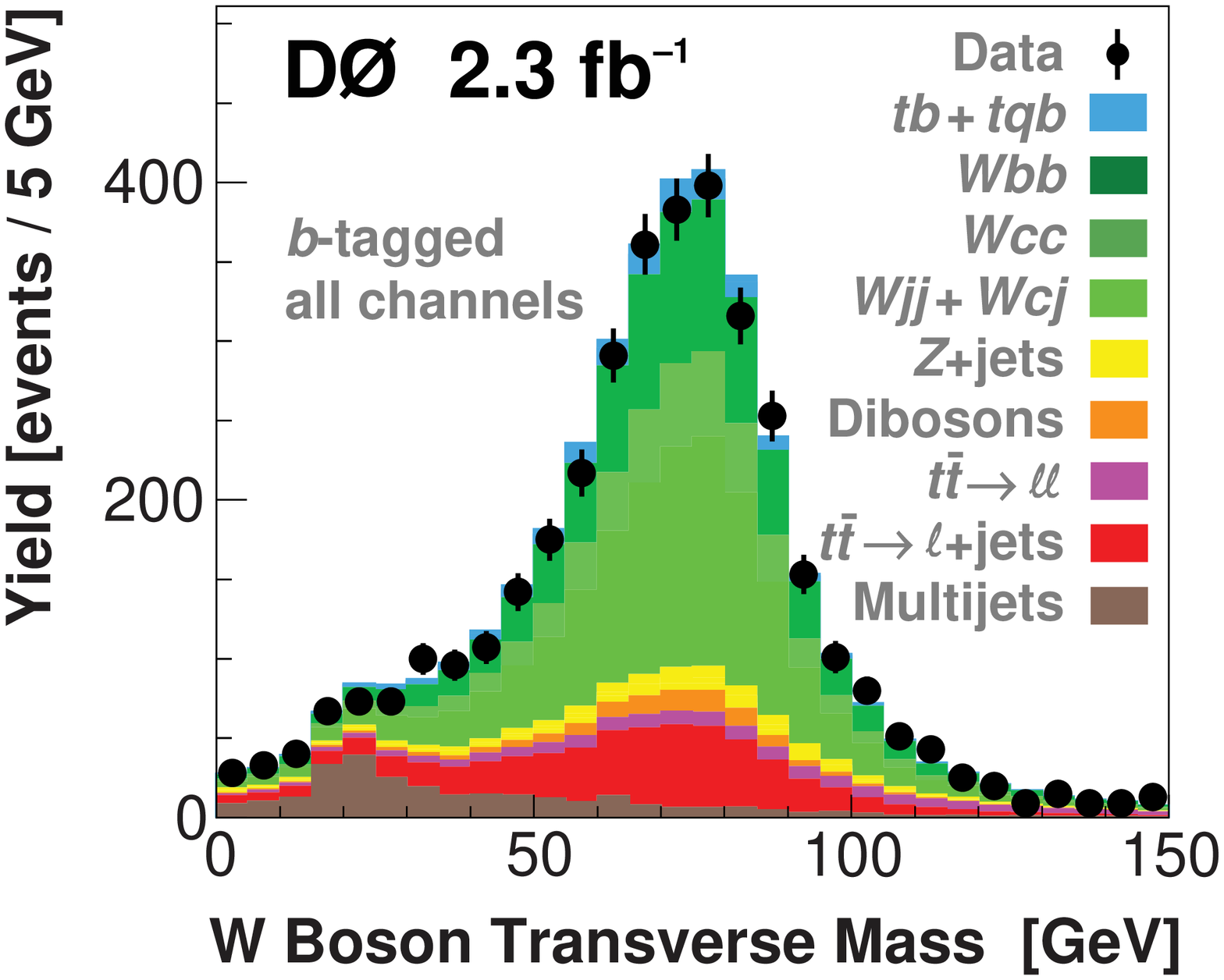}
\vspace{0.3in}\\
\includegraphics[width=2.5in]{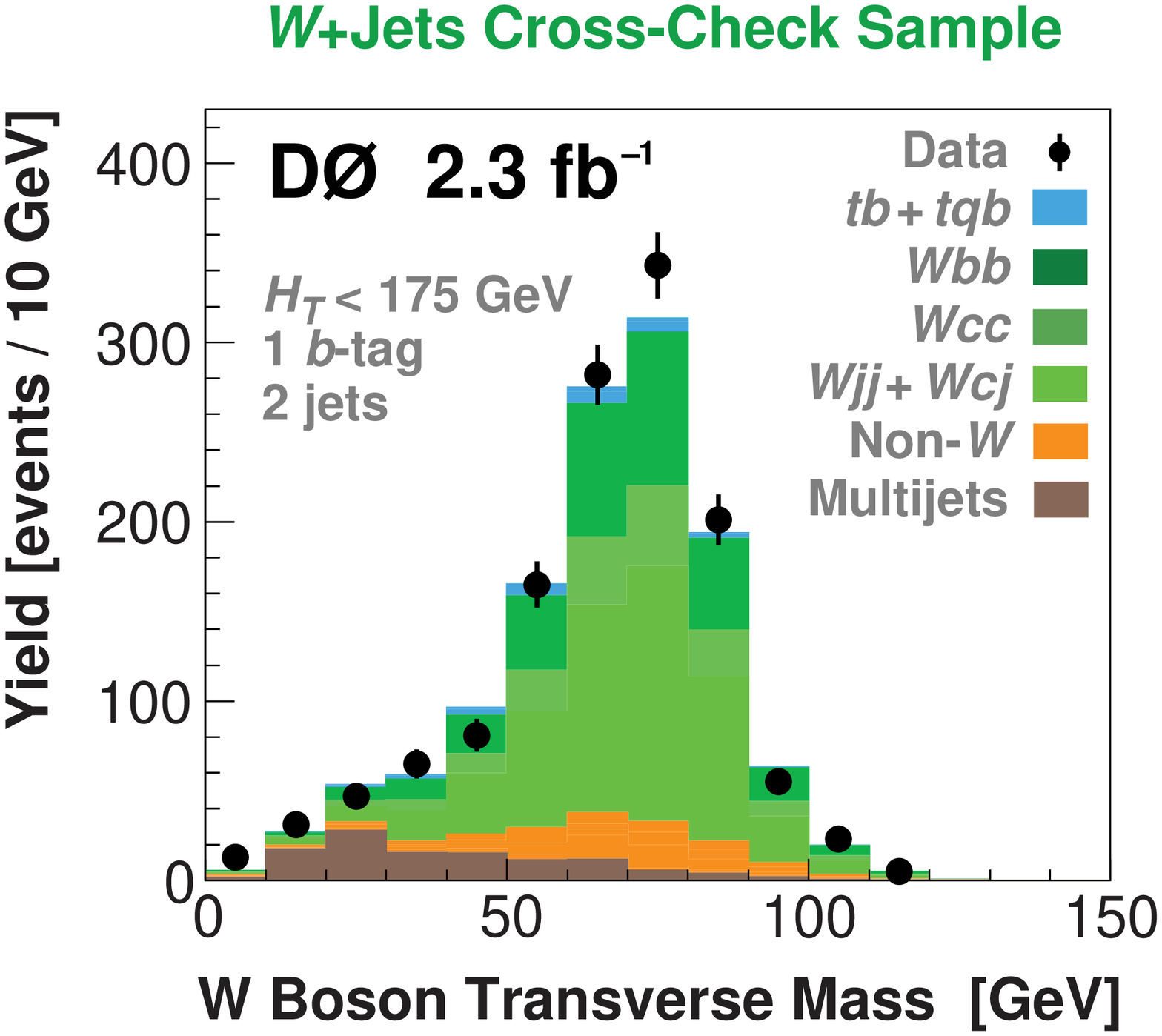}
\includegraphics[width=2.5in]{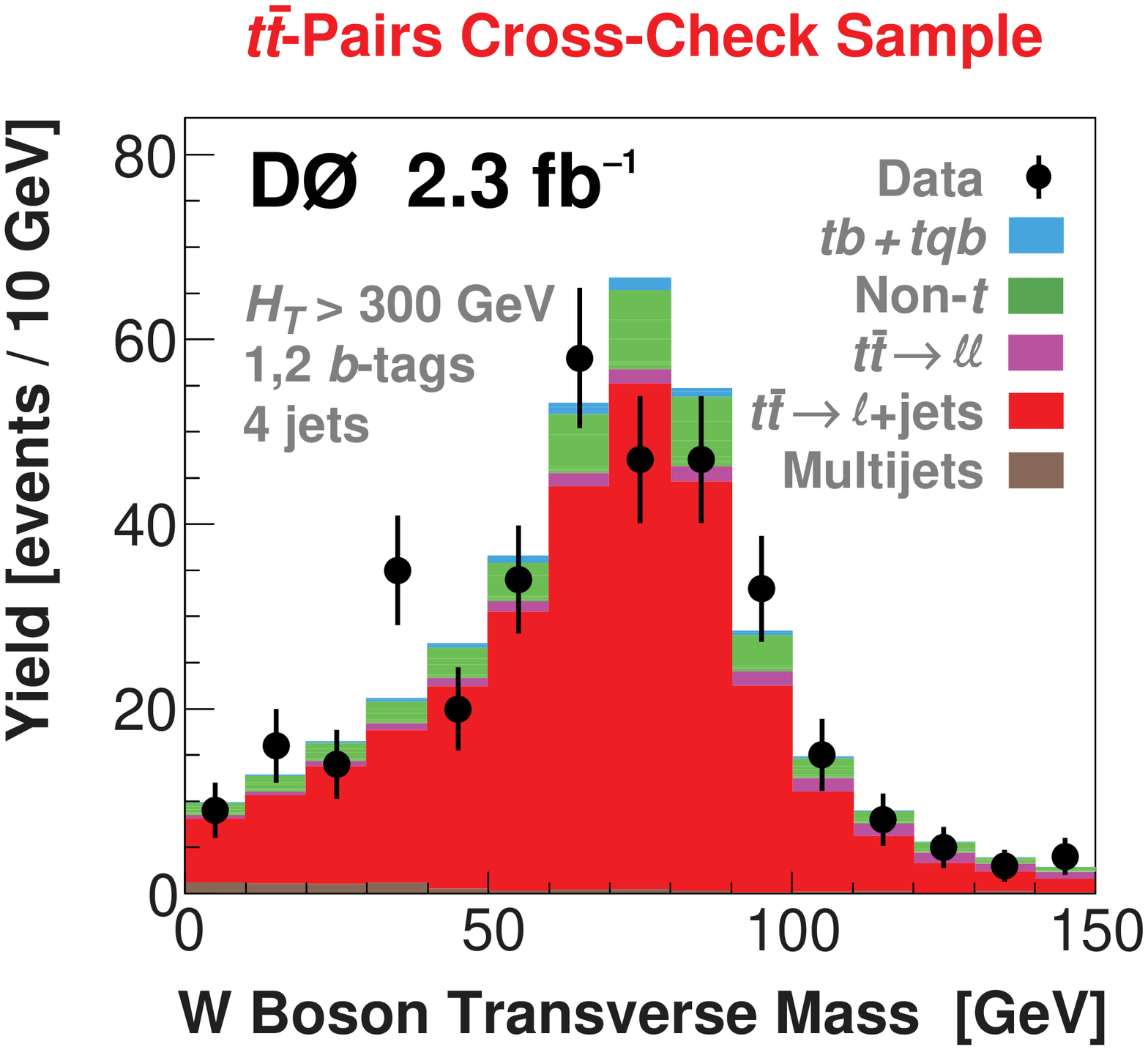}
\caption{$W$ boson transverse mass distribution shows good agreement for both 
normalization and shape of the background model for our signal sample before tagging 
(top left), after tagging (top right), and for our $W$+jets (bottom left) and
\ttbar\ (bottom right) cross-check samples.} 
\label{fig:MTW}
\end{figure*}

The background model has been checked for normalization and shape 
in hundreds of distributions in each of the 
24 individual analysis channels, before and after tagging. 
In addition, 
we also define two cross-check samples to check the background model components 
separately for the two main backgrounds: $W$+jets, and \ttbar~. 
The $W$+jets dominated sample has low $H_T$, 
exactly two jets,  
and only one $b$-tagged jet. The \ttbar\ dominated sample sample has high $H_T$, 
exactly four jets, and one or two $b$-tags. 
We find good agreement for both normalization and shape in all variables studied for 
our signal and cross-check samples. Figure~\ref{fig:MTW} shows the
$W$ transverse mass distribution for all 24 channels combined 
as an example of such tests.   

%%%%%%%%%%%%%%%%%%%%%%%%%%%%%%%%%%
\section{Multivariate Analysis Techniques}
As can be seen in Table~\ref{tab:event-yields}, the uncertainty on the background 
is larger than the expected signal. We therefore need to improve the discrimination 
between signal and background employing multivariate
analysis (MVA) techniques . 

We have improved and optimized three of these techniques following our 
2006 single top evidence analysis, described in detail in 
Ref.~\cite{d0-prd-2008}: 
boosted decision trees
(BDT)~\cite{decision-trees,bdt-benitez,bdt-gillberg}, Bayesian neural
networks (BNN)~\cite{bayesianNNs,bnn-tanasijczuk}, and the matrix
element (ME) method~\cite{matrix-elements,me-pangilinan}. 
Improvements include a larger set of input variables for the BDT and BNN analysis
from five categories: single object kinematics, global event kinematics, jet 
reconstruction, top quark reconstruction, and angular correlations. 
The BDT method uses a common set of 
64 variables for all analysis channels, while the BNN method
uses the RuleFitJF algorithm~\cite{rulefit} to select the most
sensitive kinematic variables, keeping between 18 and 28 of these as inputs, depending
on the analysis channel. The ME analysis
uses only 2-jet and 3-jet events and splits the analysis into
low and high $H_T$ regions. An additional improvement is the inclusion of 
matrix elements for more background sources: \ttbar\ , $WW$,
$WZ$, and $ggg$ diagrams in the 2-jet bin and $Wugg$ in the 3-jet bin. 

\begin{figure*}[!h!tbp]
\centering
\includegraphics[width=2.5in]{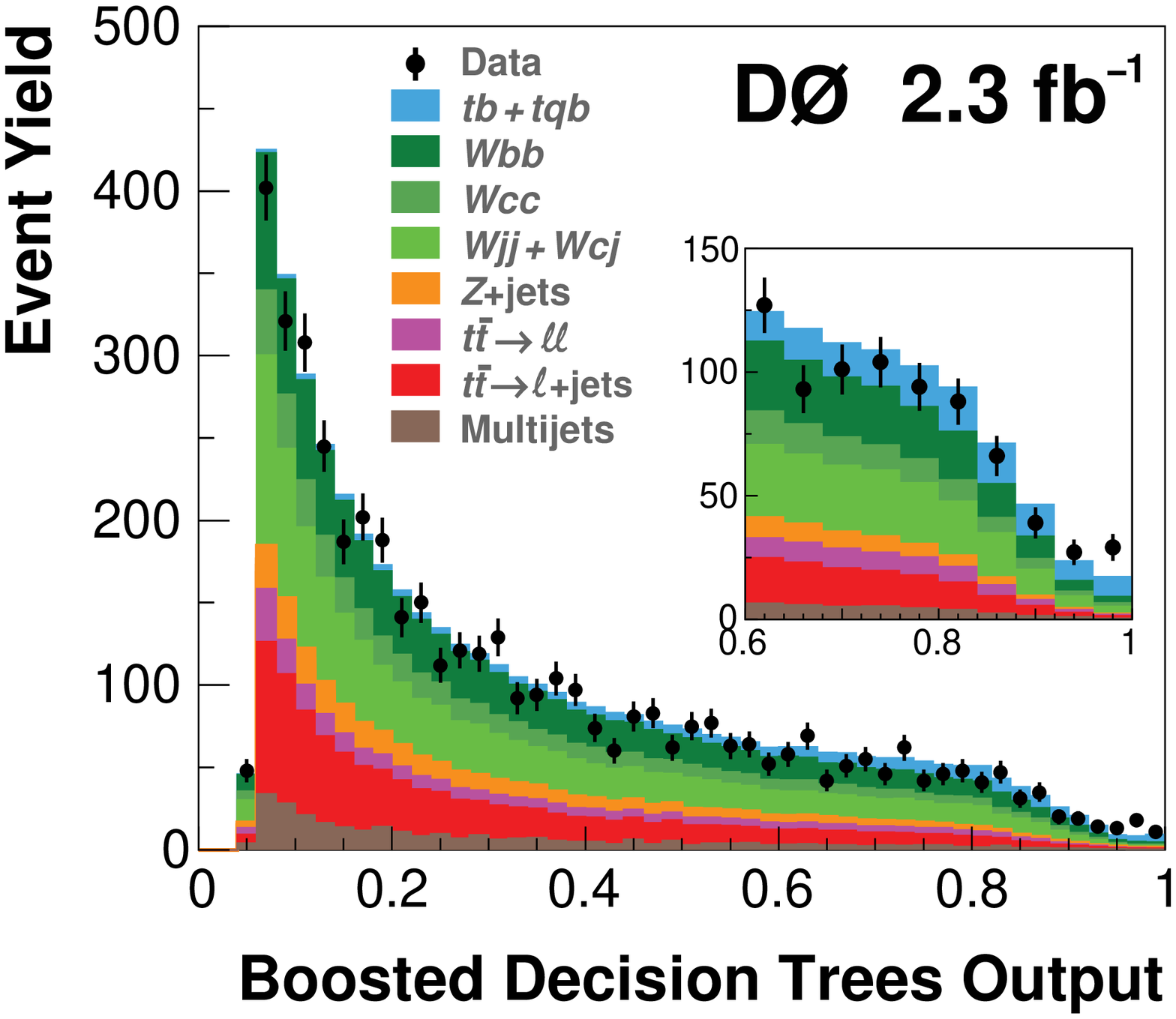}
\includegraphics[width=2.5in]{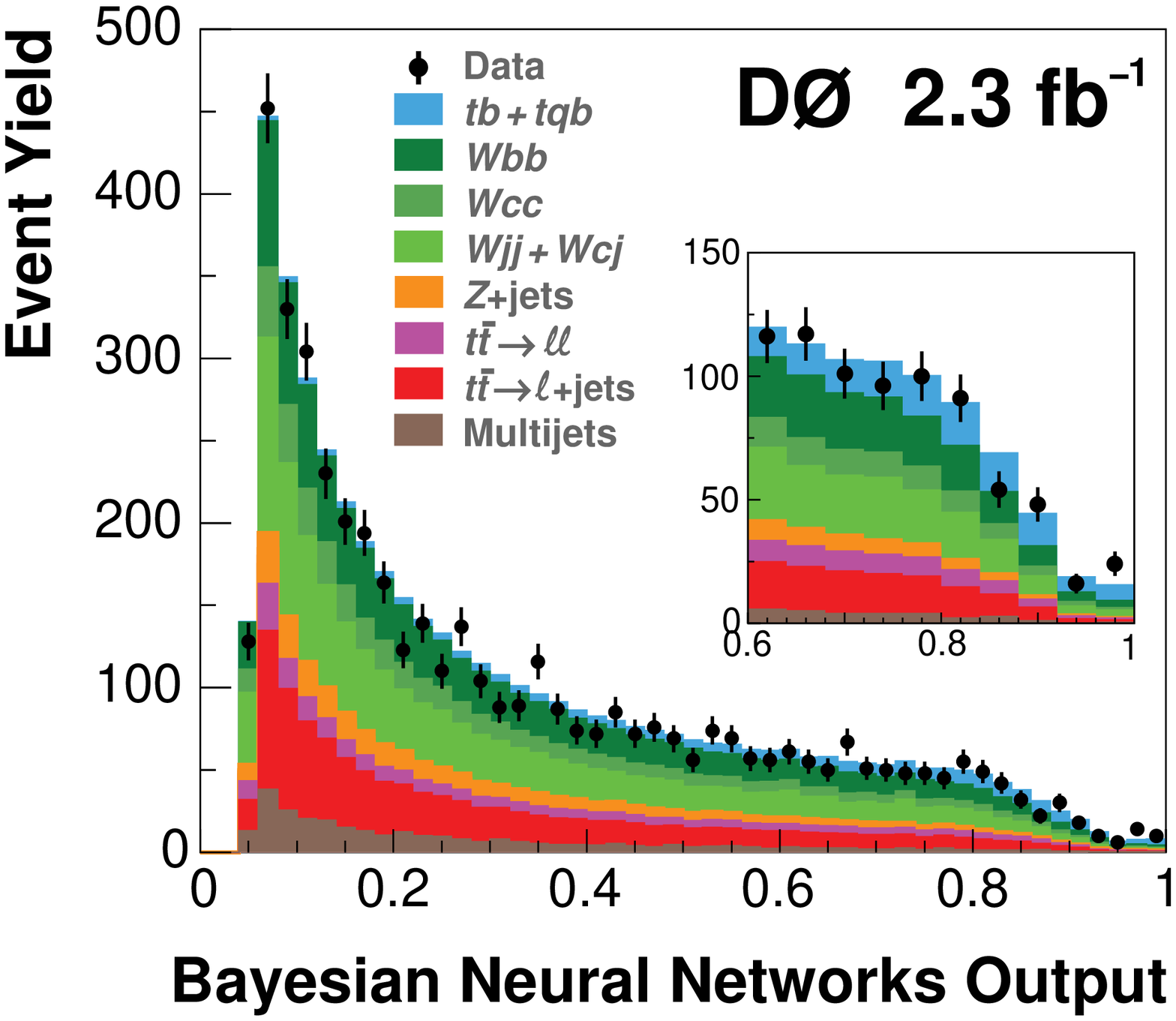}
\vspace{0.3in}\\
\includegraphics[width=2.5in]{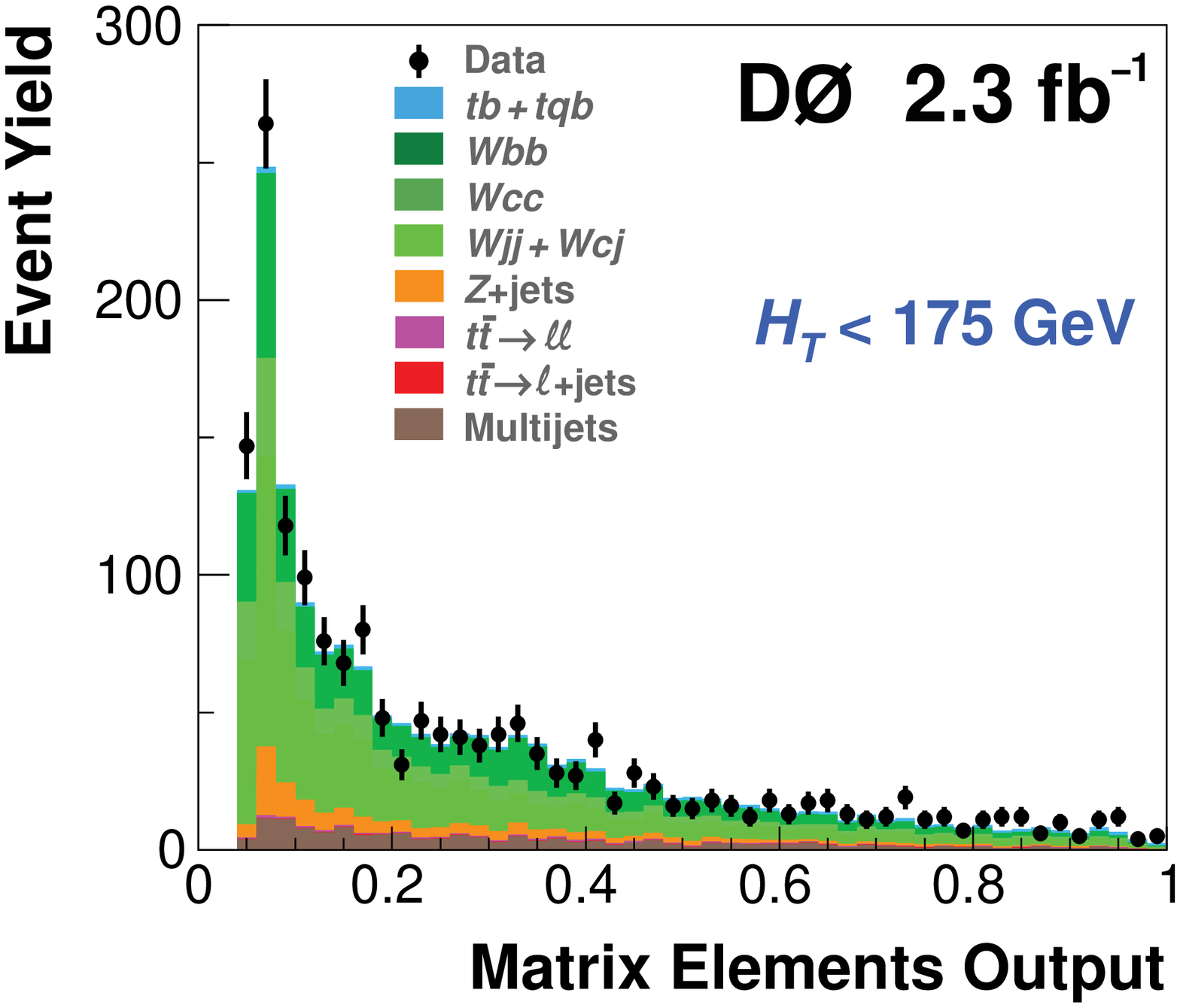}
\includegraphics[width=2.5in]{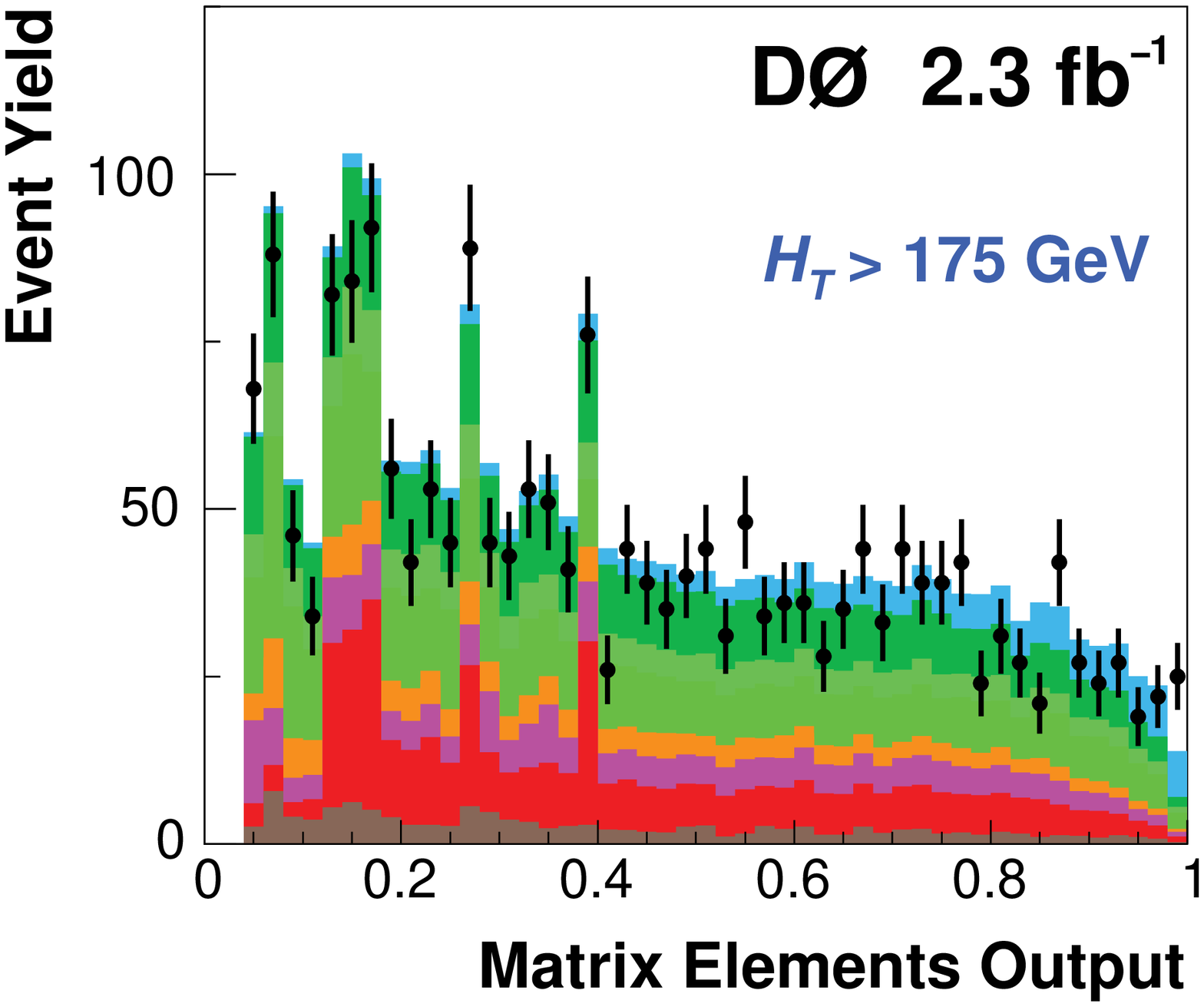}
\caption{Discriminant
output distributions for BDT (top left), BNN (top right), ME low $H_T$ (bottom left), 
and ME high $H_T$ (bottom right). The bins from the matrix elements outputs are 
reordered in descending S:B from 1 toward 0. Analysis channels with 
lower statistics do not have entries in all 50 bins, which creates the spikes 
that can be seen in the plots.} 
\label{fig:discrim}
\end{figure*}

All three analyses transform their
output distributions to ensure that every bin has at least 40 background MC 
events before normalization, 
so that there are no bins with a nonzero signal prediction or data but 
not enough background in the model to use that information. Figure~\ref{fig:discrim} 
shows the discriminant output distributions for the three MVA techniques. It is 
important to note that the plots show the sum of the individual discriminant outputs 
for each of the 24 individual analysis channels, which are treated as independent in 
the cross section extraction.  

Even though the three MVA techniques use the same data sample they are not 100\%
correlated, in fact, BDT and BNN are $\approx$~75\% correlated with each other and
$\approx$~60\% correlated with ME. We therefore combine these methods using 
an additional BNN (BNNComb) that takes
as input the output discriminats of the BDT, BNN and ME methods, and produces a single
combination output discriminant. This BNNComb leads to an increased
expected sensitivity and a more precise measurement of the single top cross section. 

\begin{figure*}[!h!tbp]
\includegraphics[width=2.5in]{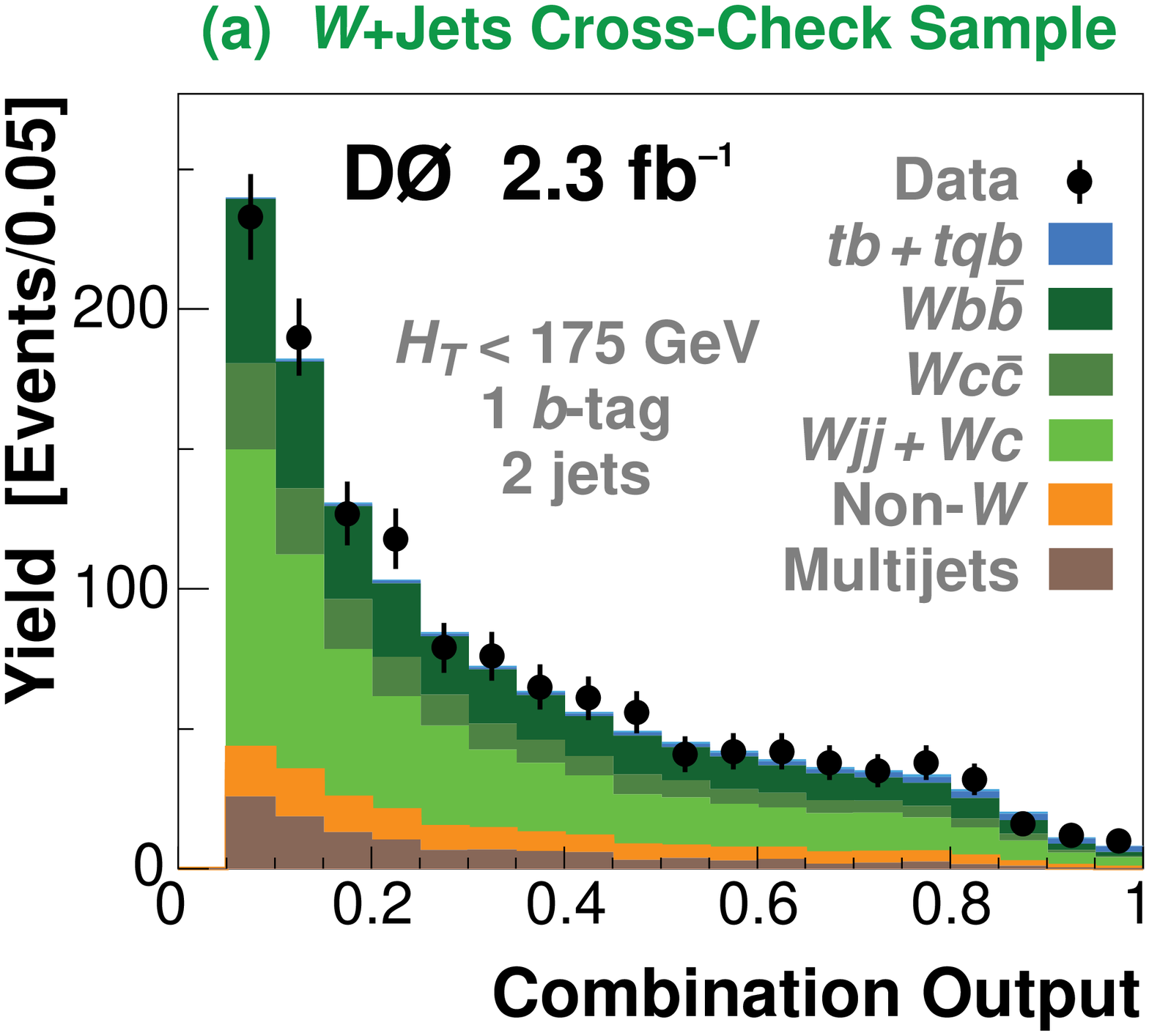}
\includegraphics[width=2.5in]{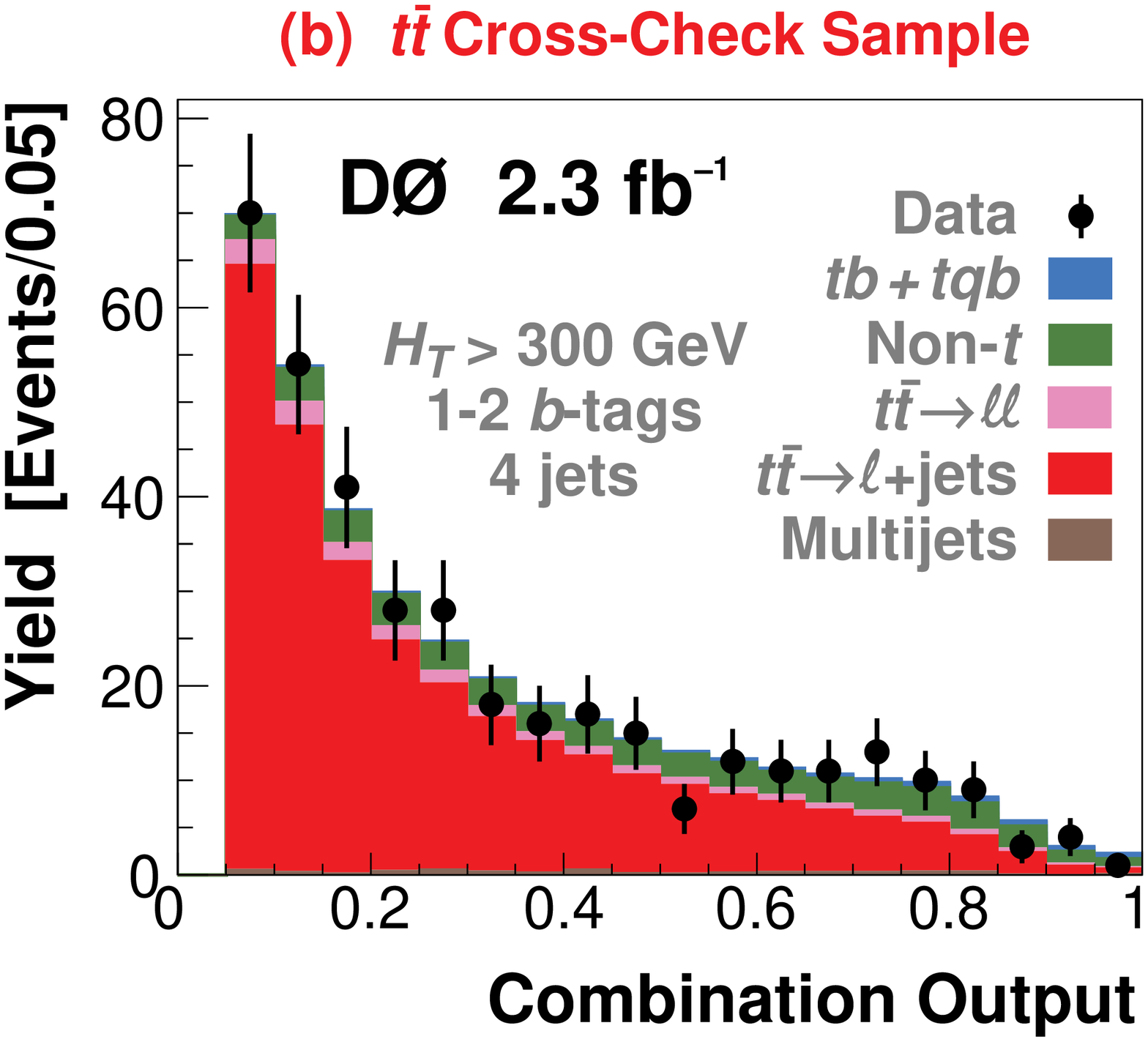}
\caption[crosschecks]{The combination discriminant outputs for
(a)~$W$+jets and (b)~{\ttbar} cross-check samples.}
\label{fig:BNNCombxcheck}
\end{figure*}

We tested each MVA technique (BDT, BNN, ME and BNNComb) generating 
ensembles of pseudodatasets created from background and signal at
different cross sections to confirm a linear response and an unbiased
cross section measurement. The output discriminants have also been produced 
for the $W$+jets and \ttbar\ cross-check samples and demonstrate that 
the backgrounds are
well-modeled across the full range of the discriminant output. 
Figure~\ref{fig:BNNCombxcheck} shows the combination discriminant outputs for the
cross-check samples.
 
\section{Cross Sections}
The single top production cross section is measured using a Bayesian 
approach as in our previous results~\cite{d0-prl-evidence, d0-prd-2008}. A 
binned likelihood is formed as a
product over all bins and channels, evaluated separately for each
MVA technique. The central value of the cross section is
given by the position of the peak in the posterior density, and the
$68\%$ interval about the peak gives the $\pm 1\;\rm standard \; deviation \; (SD)$ 
uncertainty. 
The sensitivity of each analysis to a contribution from single top
quark production is estimated by generating ensembles of 
pseudodatasets that sample the background model and its uncertainties
in the absence of signal. A cross section is measured from each
pseudodataset, which allows us to calculate the probability to measure the SM cross
section (``expected significance'') or the observed cross section 
(``observed significance''). 
Table~\ref{tab:xsec} summarizes the measured cross section, and the expected and observed
significance for each MVA technique. The cross section measured by the BNNComb has a 
$p$-value of $2.5\times10^{-7}$,
corresponding to a significance of $5.0\;\rm SD$, and to the observation of 
single top production.

\begin{table}[!h!btp]
\caption{Measured cross section, and the expected and observed
significance for each MVA technique.}
\label{tab:xsec}
\begin{tabular}{|c|c|c|c|}\hline
MVA & $\sigma \pm \Delta\sigma$~(pb) & Expected (SD)& Observed (SD) \\ 
\hline
BDT         &  $3.74\pm^{0.95}_{0.79}$ & 4.3 & 4.6 \\
BNN         &  $4.70\pm^{1.18}_{0.93}$ & 4.1 & 5.2 \\
ME     	    &  $4.30\pm^{0.99}_{1.20}$ & 4.1 & 4.9 \\ \hline
BNNComb     &  $3.94 \pm 0.88$ & 4.5 & 5.0 \\   \hline
\end{tabular}
\end{table}

\begin{figure*}[!h!tbp]
\includegraphics[width=2.5in]{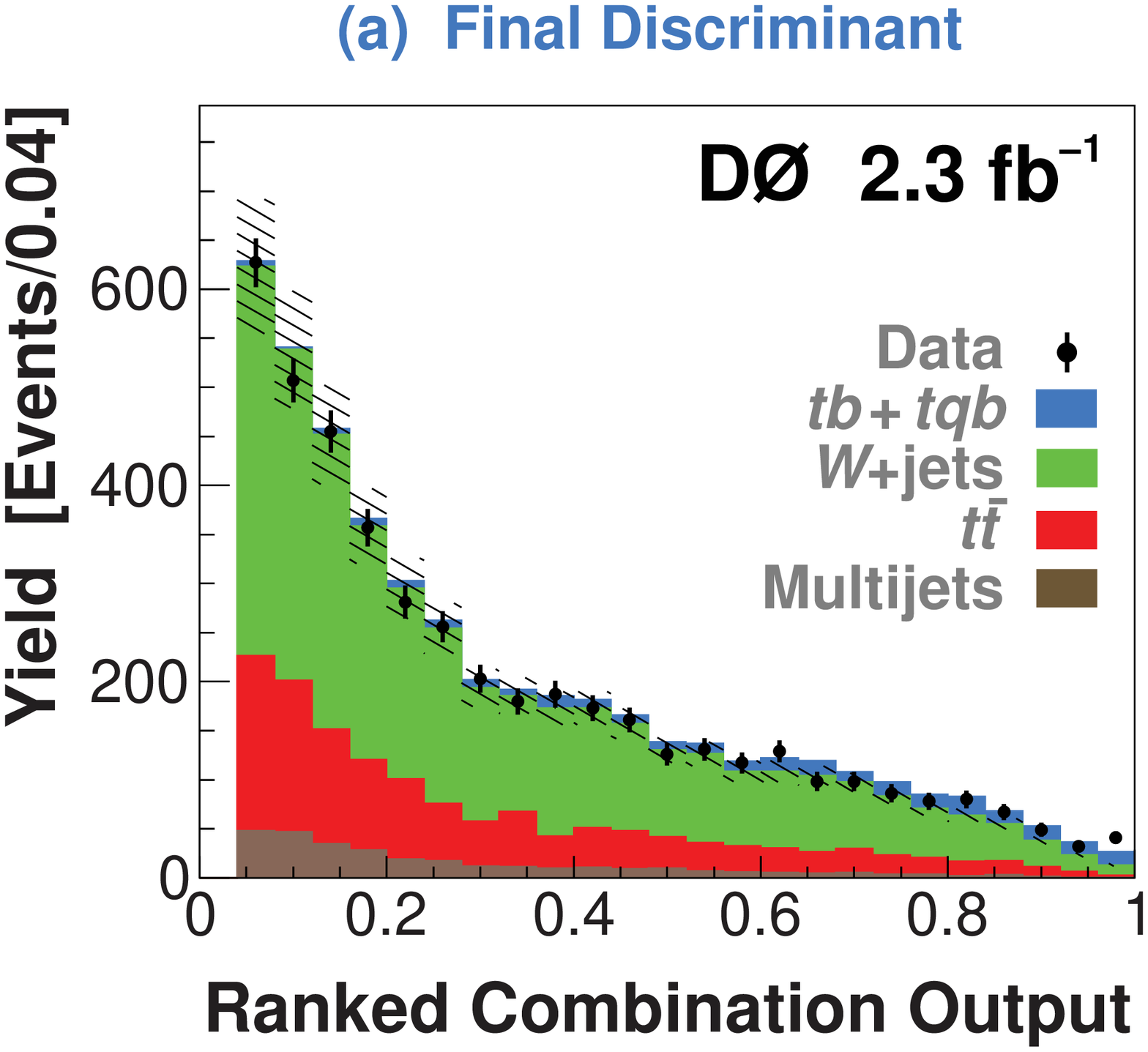}
\includegraphics[width=2.5in]{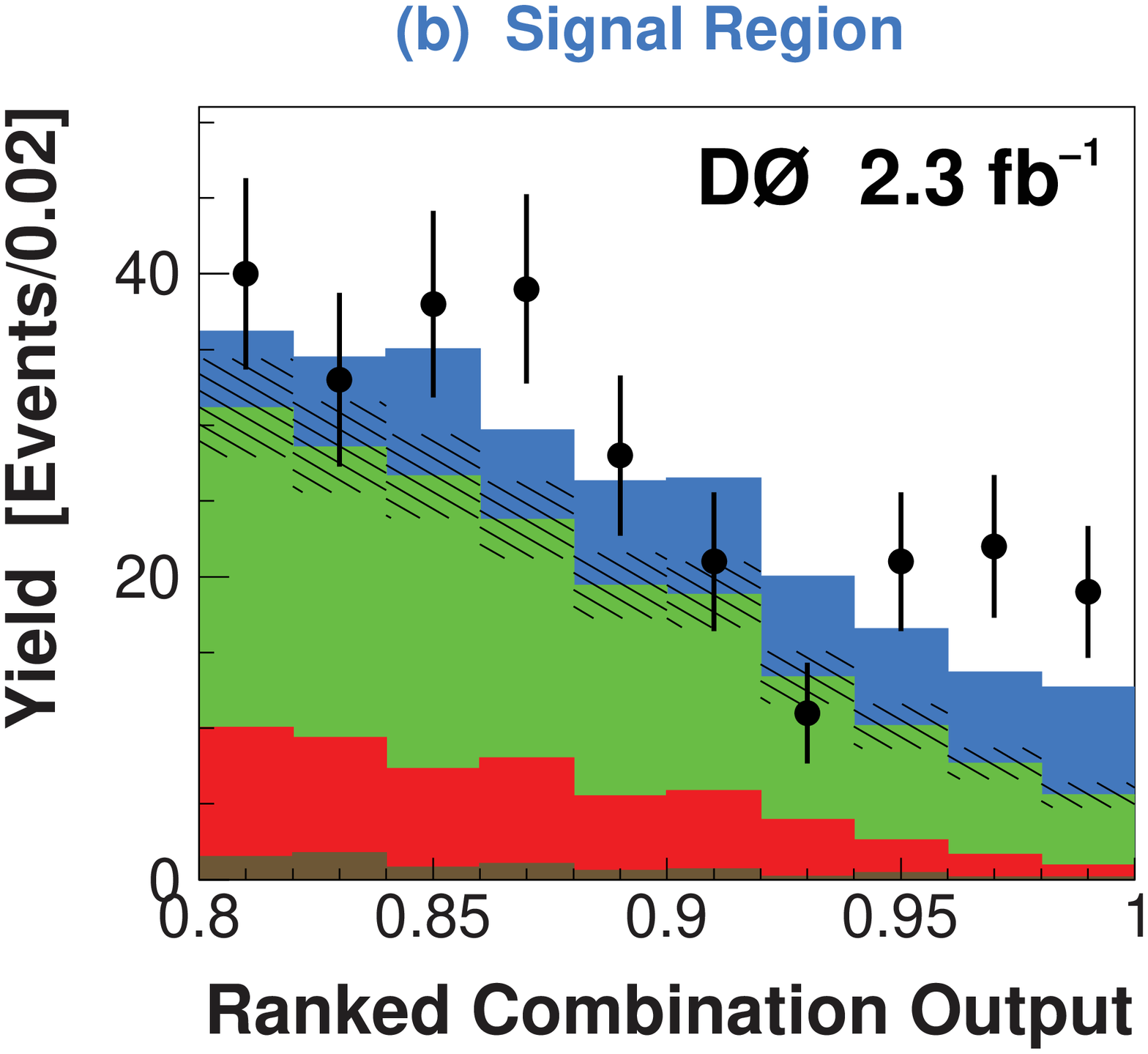}\\
\vspace{0.1in}
\includegraphics[width=2.5in]{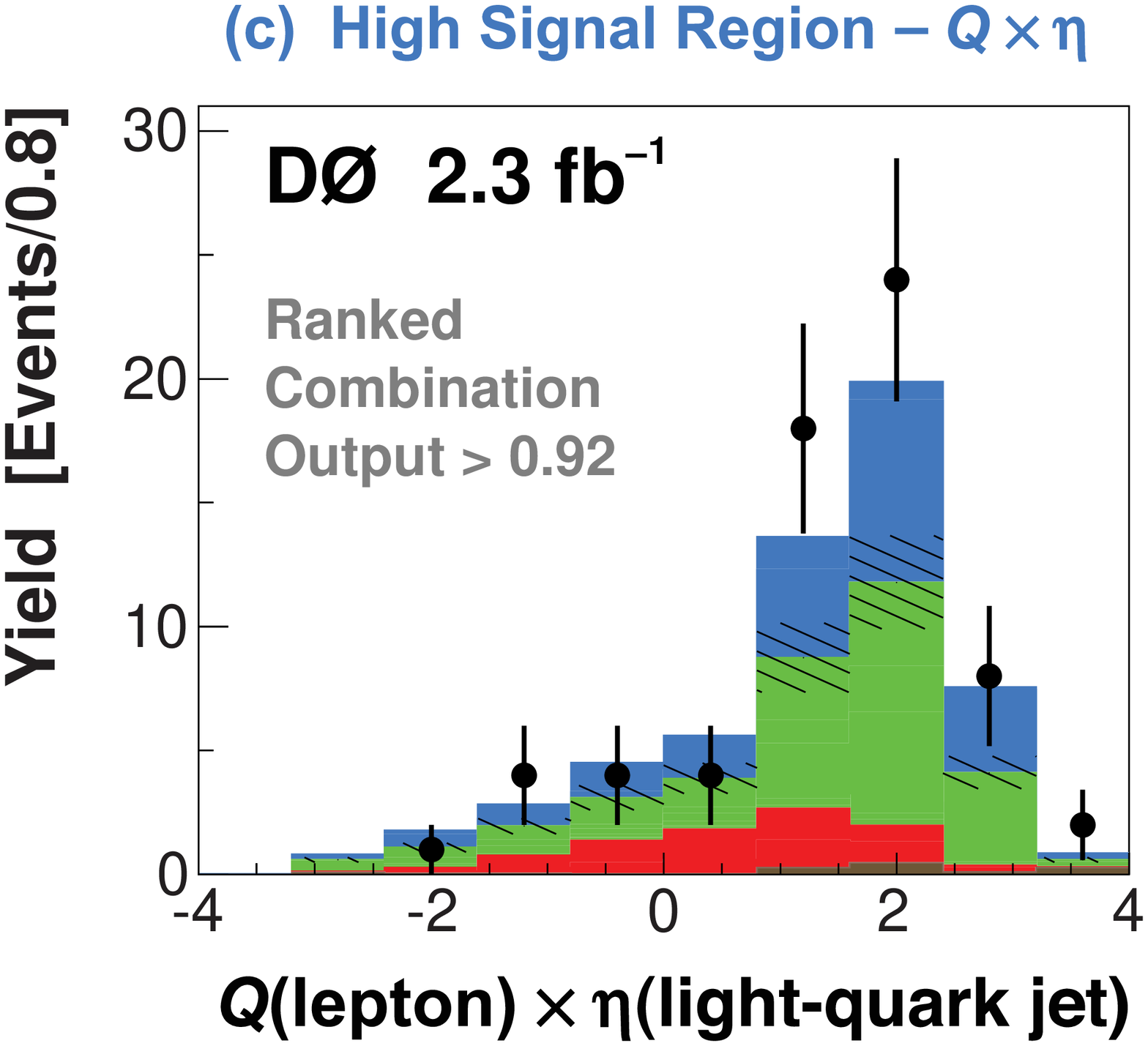}
\includegraphics[width=2.5in]{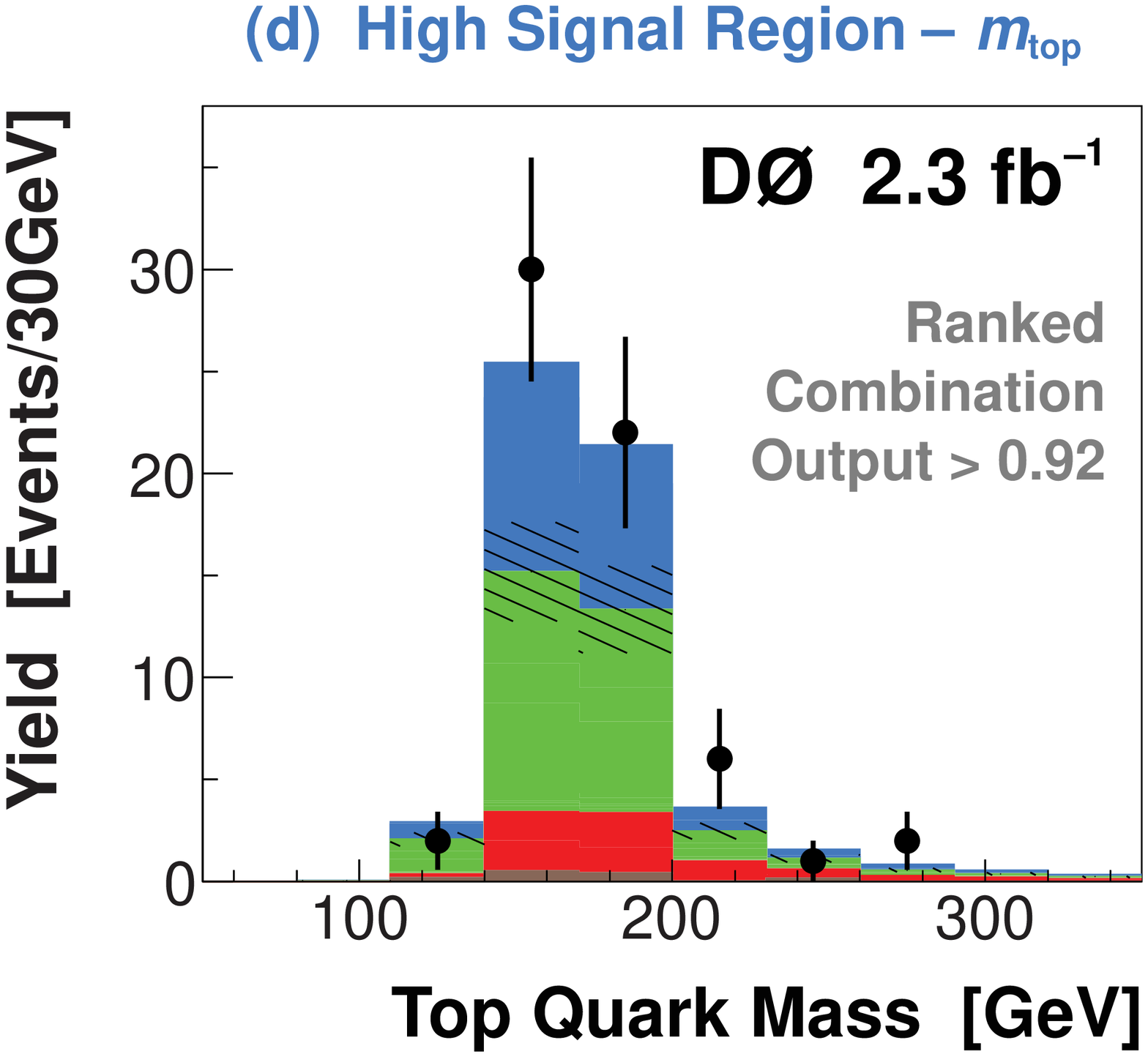}\\
\vspace{-0.1in}
\caption[discriminants]{Distribution of the combination output for the 
full range~(a), and the high signal
region~(b). The bins have been ordered by their expected S:B
ratio and the signal is normalized to the measured cross section. The
hatched band indicates the total uncertainty on the background. 
Distribution of lepton charge times pseudorapidity of
the leading not-$b$-tagged jet~(c), and reconstructed top
quark mass~(d) for events with ranked combination output $> 0.92$. 
}
\label{fig:results}
\end{figure*}

\begin{figure}[!h!tbp]
\includegraphics[width=3in]{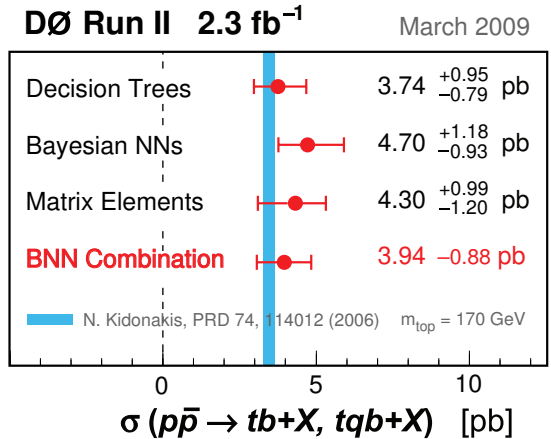}
\caption[discriminants]{Summary of the measured cross sections compared to
theoretical predictions. All measurements are in good agreement with each other and
with the SM prediction. 
}
\label{fig:xsecsum}
\end{figure}

Figure~\ref{fig:results} shows the distribution of the combination output and
examples of variables with high sensitivity to the signal, which 
illustrate the importance of the signal to achieve a good modeling of the 
data. Figure~\ref{fig:xsecsum} summarizes the cross sections measured by 
each of the MVA techniques and compares them to theoretical 
predictions~\cite{singletop-xsec-kidonakis}. 

\section{$t$-Channel Production}
The 
analysis presented in the previous section measured the combined single top cross 
production section assuming the ratio between the $s$- and $t$-channel production modes
to be fixed, and given by the SM. However,  
this ratio could be modified by beyond-the-SM physics processes, for instance 
additional quark generations,
new heavy bosons~\cite{Tait:2000sh}, flavor-changing neutral currents~\cite{d0-fcnc}, or
anomalous top quark couplings~\cite{singletop-wtb-heinson,d0-singletop-wtb,Abazov:2009ky}.
It is therefore of interest to measure the two production modes individually, and 
compare them to theoretical predictions.  

\begin{figure}[!h!tbp]
\includegraphics[width=2.5in]{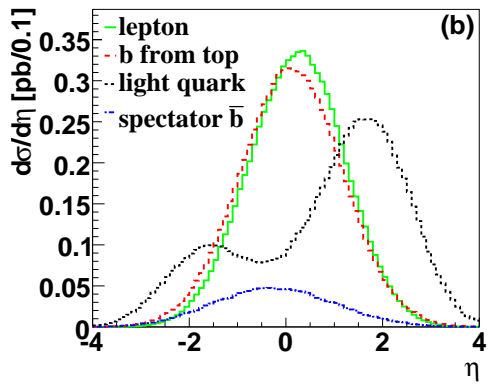}
\caption[discriminants]{Parton-level pseudorapidity distribution of the final
state objects in top production. The pseudorapidity distribution of the 
light quark jet is the most sensitive variable to 
separate $t$-channel events from both the $s$-channel and the
backgrounds.}
\label{fig:tchannel}
\end{figure}

In this section I present a recent D0 analysis~\cite{tchannel} 
that uses the same dataset, event
selection, and signal/background modeling as the observation analysis presented
in the previous section, but the MVA filters are specifically trained to extract 
the single top quark events produced via the $t$-channel mode. This results in a 
measurement of the $t$-channel cross section that is independent of the $s$-channel 
cross section model. Even though we use the same 
set of variables as in the observation
analysis, only $t$-channel single top events are considered signal during the
optimization. $s$-channel single top events are considered together with other 
background processes, and its rate is 
normalized to the SM expectation~\cite{singletop-xsec-kidonakis}. 
Figure~\ref{fig:tchannel} 
shows the parton-level pseudorapidity distribution of the final
state objects in top production (excluding antitop). 
The most sensitive variable that
separates $t$-channel events from both the $s$-channel and the
backgrounds is the pseudorapidity distribution of the 
light quark jet. 

We use the same three MVA techniques (BDT, BNN and ME) and a forth one for the combination 
(BNNComb). We test each of them for linearity and unbiased cross section 
measurement, and confirm that the backgrounds are well-modeled across the 
entire range of the output discriminants for the $W$+jets and 
\ttbar\ cross-check samples. 

\begin{figure}[!h!btp]
\includegraphics[width=2.5in]{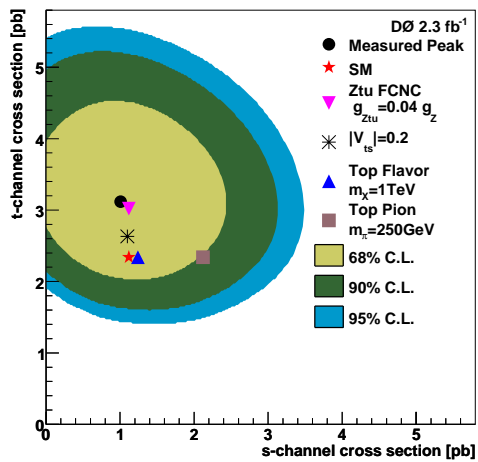} 
\caption{Posterior probability density for $t$-channel and $s$-channel single 
top quark production in contours of equal probability density. 
The various points correspond to the the measured cross section, the SM
expectation, and a sample of beyond-the-SM predictions.}
\label{fig:posterior_2D}
\end{figure}

We use the same Bayesian statistical analysis to measure the
production cross sections. First, we compute the two-dimensional
posterior probability density as a function of both $t$-channel and $s$-channel 
single top cross sections. The resulting posterior probability density is 
shown in Fig.~\ref{fig:posterior_2D}. 

Next, we integrate over the $s$-channel axis in Fig.~\ref{fig:posterior_2D} to 
obtain the $t$-channel posterior probability density without any 
assumptions about the value of the $s$-channel cross section. 
We have verified the linearity of the procedure and its 
independence of the input $s$-channel cross section with ensembles
of pseudo-datasets generated at several different $t$-channel and $s$-channel 
cross sections. An equivalent procedure is performed to 
obtain the $s$-channel posterior probability density by 
integrating over the $t$-channel axis.
The resulting production cross section for 
single top quark production is $3.14^{+0.94}_{-0.80}$~pb for the $t$-channel and 
$1.05 \pm 0.81$~pb for the $s$-channel. Both are 
in good agreement with the SM predictions~\cite{singletop-xsec-kidonakis} 
of $2.34 \pm 0.13$~pb and $1.12 \pm 0.04$~pb, respectively (for a top quark mass of 
$m_t=170$~GeV). 
The measured $t$-channel cross section has a 
p-value of  $8.0\times 10^{-7}$,
corresponding to a Gaussian significance of 4.8$\sigma$, and represents the first 
direct evidence of single top quark $t$-channel production. 

\section{Summary}
The D0 collaboration has reported the observation of 
single top quark production using 2.3~fb$^{-1}$ 
of data collected at the Fermilab Tevatron. 
The measured cross section for the combined $tb$+$tqb$ channels is 
$\sigma(p\overline{p} \to tb+X,~tqb+X) = 3.94 \pm 0.88$~pb, corresponding to an 
excess of signal over the predicted background with a significance of
$5.0\,\sigma$. This measurement assumes the ratio between the 
$s$- and $t$-channel production modes
is fixed, and given by the SM. Using the same dataset and MVA techniques, the analysis 
is modified to remove this assumption and uses the $t$-channel characteristics 
to measure the $t$-channel and $s$-channel cross sections simultaneously. This results
in a $t$-channel measurement that is independent of the $s$-channel cross section model.
The resulting cross sections are 
$\sigma(p\overline{p} \to tqb+X)=3.14^{+0.94}_{-0.80}\;\rm pb$ and 
$\sigma(p\overline{p} \to tb+X)=1.05 \pm 0.81\;\rm pb$, in agreement with SM 
predictions. The measured $t$-channel cross section corresponds to an 
excess of signal over the predicted background with a significance of
$4.8\,\sigma$, and is the first
analysis to isolate an individual single top quark production channel.

%%%%%%%%%%%%%%%%%%%%%%%%%%%%%%%%%%
\begin{acknowledgments}
We thank the staffs at Fermilab and collaborating institutions, 
and acknowledge support from the 
DOE and NSF (USA);
CEA and CNRS/IN2P3 (France);
FASI, Rosatom and RFBR (Russia);
CNPq, FAPERJ, FAPESP and FUNDUNESP (Brazil);
DAE and DST (India);
Colciencias (Colombia);
CONACyT (Mexico);
KRF and KOSEF (Korea);
CONICET and UBACyT (Argentina);
FOM (The Netherlands);
STFC (United Kingdom);
MSMT and GACR (Czech Republic);
CRC Program, CFI, NSERC and WestGrid Project (Canada);
BMBF and DFG (Germany);
SFI (Ireland);
The Swedish Research Council (Sweden);
CAS and CNSF (China);
and the
Alexander von Humboldt Foundation (Germany).
\end{acknowledgments}

\bigskip % extra skip inserted

\end{document}